\documentstyle[aps,multicol,epsfig]{revtex}

\begin{document}

\title{Absence of surface mode in a visco-elastic material with
surface tension}

\author{ Hiizu Nakanishi and Satoshi Kubota$^*$}
\address{ Department of Physics, Kyushu University 33,
Fukuoka 812-8581, Japan}

\date{}

\maketitle

\begin{abstract}
The surface waves in the visco-elastic media with the surface tension
are studied using the Voigt-Kelvin model of the visco-elasticity.
It is shown that the surface mode of oscillation does not exist in the 
parameter region where the effect of surface tension is larger than
that of the elastic stress at the surface unless the viscous stress
masks the elastic stress in the bulk.
In the region, the surface oscillation is suppressed and the oscillation
beneath the surface diffuses after the pulse goes into the bulk.
The experimental relevance of the present results is also discussed.
\end{abstract}



\begin{multicols}{2}


Surface modes of oscillation in a material are quite different from bulk
modes due to the boundary condition at the surface.
In an elastic material, the longitudinal and the transverse modes are
mixed in the Rayleigh mode (surface mode), and its speed is slower
than both of them.
As for the fluid where only restoring force is the surface tension,
the dispersion of the surface wave is not linear but given by
$\omega\propto k^{3/2}$.
The transition between the two modes in the visco-elastic material
like polymer solutions or gels has 
been studied theoretically\cite{php88,hpp91,jk95}
and experimentally\cite{kst91,kst94,dt93,ckc95,tc96,mohm94}.

The mode characters of these two modes, however, are quite different and
they cannot mix easily in a visco-elastic material where both the
elastic and the surface tension operate.
It has been expected that the spectral peak of the surface tension
wave in the thermal fluctuation does not transform into the Rayleigh
peak smoothly but the two peaks tend to co-exist in the transition
region, and the spectral structure does not be simple\cite{hpp91}.
The expected shoulder structure of the spectral peak has not been
observed clearly but the
substantial increase of the increase of the spectral width was
demonstrated in the transition region for polymer solutions upon
increasing density\cite{dt93,ckc95}

Recently, a careful experiment was done to measure the speed and
the attenuation of the {\em externally excited} surface wave
in the sol and the gel phase of tungstic acid\cite{okhk97},
which undergoes a sol-gel transition upon decreasing $pH$,
and it was shown that
there is an anomalous behavior around the sol-gel transition point; 
the wave speed rises sharply and then drops discontinuously,
and the attenuation rate has a sharp peak.
These behaviors have not been expected by the previous theories.


Being motivated by the above experiment, we analyze the dispersion
relation of surface modes of a visco-elastic material.  We focus our
attention to the experimental situation where the inverse of the
surface wave frequency is much smaller than the structural relaxation
time of the polymer network over the wave length.  Then the Maxwell
type of visco-elasticity can be ignored and the Voigt-Kelvin
visco-elasticity, in which the stress is the sum of the elastic and
the viscous stresses, is a good model.
Note that the Voigt-Kelvin model does not become particularly bad
around the sol-gel transition point because we are interested in the
phenomenon with finite frequency and wave length.
The sol phase behavior is also obtained within this model by setting
the elasticity zero.

With this model, we demonstrate that the surface mode does not exist around
the region where the surface tension competes with the elastic stress.


We assume the visco-elastic material is occupied in the $z>0$ space and
the surface is located at the $x-y$ plane.
The displacement field $\vec u(x,y,z,t)$ can be decomposed into the
transverse part $\vec u_t$ and the longitudinal part $\vec u_l$ as
\begin{equation}
\vec u=\vec u_t + \vec u_l;
\qquad
\vec\nabla\cdot\vec u_t=0,\quad
\vec\nabla\times\vec u_l=0 ,
\end{equation}
then the equations of motion for $\vec u_t$ and $\vec u_l$
can be expressed as
\begin{eqnarray}
\rho\ddot{\vec u_t} & = &
\left( E+\eta{\partial\over\partial t}\right)
\nabla^2\vec{u_t} \, ,
\label{eq.of.m.t} \\
\rho\ddot{\vec u_l} & = &
\left[ 
{4\over 3}\left( E+\eta{\partial\over\partial t}\right)
+
\left( K+\zeta{\partial\over\partial t}\right)
\right]
\nabla^2\vec{u_l}\, ,
\label{eq.of.m.l}
\end{eqnarray}
where the dot denotes the time derivative and $\rho$ is the mass
density, $E$ and $K$ are the shear and bulk modulus, and $\eta$ and
$\zeta$ are the shear and bulk viscosity, respectively.
Using a potential function $\phi_l$,
the longitudinal displacement $\vec u_l$ can be expressed as
\begin{equation}
\vec u_l = \vec\nabla\phi_l,
\label{l.comp}
\end{equation}
then equation of motion (\ref{eq.of.m.l}) is expressed as
\begin{equation}
\rho\ddot\phi_l =
\left[ 
{4\over 3}\left( E+\eta{\partial\over\partial t}\right)
+
\left( K+\zeta{\partial\over\partial t}\right)
\right]
\nabla^2\phi_l .
\end{equation}
The stress tensor $\sigma_{ij}$ is given by
\begin{eqnarray}
\sigma_{ij} & = &
2\left(E+\eta{\partial\over\partial t}\right)
\left(\varepsilon_{ij}-{1\over 3}{\rm Tr}[\varepsilon]\,
\delta_{ij}\right)+
\nonumber \\ 
& & \qquad
\left(K+\zeta{\partial\over\partial t}\right)
{\rm Tr}[\varepsilon] \delta_{ij},
\end{eqnarray}
where $\varepsilon$ is the strain tensor
\begin{equation}
\varepsilon_{ij}={1\over 2} \left(
{\partial u_i\over\partial x_j} + {\partial u_j\over\partial x_i}
\right).
\end{equation}

For simplicity, we take the incompressible limit, $K\to\infty$, which
is good for the experimental situation where the surface wave speed
under investigation is much slower than the longitudinal wave speed.
Then we have
\begin{equation}
\nabla^2\phi_l = 0
\end{equation}
and
\begin{equation}
\sigma_{ij}=
2\left(E+\eta{\partial\over\partial t}\right)
\varepsilon_{ij}
+
\rho\ddot\phi_l \,\delta_{ij}.
\end{equation}

Now we examine the solution of the form
\begin{eqnarray}
\vec u_t(x,z,t) & = &
\Bigl(  -f'(z,t)/ik,\, 0,\,  f(z,t) \Bigr) e^{ikx},
\label{pw.sol}
\\
\phi_l(x,z,t) & = & \phi(t)e^{ikx-kz},
\label{pw.sol.phi}
\end{eqnarray}
assuming the physical quantities does not depend on the $y$ co-ordinate
and the $y$ component of the displacement is zero.
In (\ref{pw.sol}), the prime denotes the derivative in $z$.
The parameter $k$ is the wave number in the $x$ direction; $k>0$
in order that the solution should be the surface mode.
Then the equation of motion (\ref{eq.of.m.t}) is given by
\begin{equation}
\rho\ddot f(z,t) =
\left(E+\eta{\partial\over\partial t}\right)
\left(-k^2+{\partial^2\over\partial z^2}\right) f(z,t).
\label{t-dep}
\end{equation}
The boundary conditions at $z=0$ are
\begin{equation}
\sigma_{xz}=0,\qquad \sigma_{zz}=
-\gamma{\partial^2 u_z\over\partial x^2},
\end{equation}
where $\gamma$ is the surface tension.
These can be represented as
\begin{eqnarray}
-\left( f''(0,t)+k^2 f(0,t)\right) + 2k^3\phi(t) & = & 0 ,
\label{BC.1}
\\
2\left(E + \eta{\partial\over\partial t}\right)
\left( f'(0,t)+k^2\phi(t)\right) + \rho\ddot \phi(t) & = &
\nonumber \\
 \gamma k^2\bigl( f(0,t) & - & k\phi(t)\bigr) .
\label{BC.2}
\end{eqnarray}
If we assume the sinusoidal time dependence,
\begin{equation}
f(z,t) = f_0 e^{-\kappa z -i\omega t},\qquad
\phi(t) = \phi_0 e^{-i\omega t} ,
\end{equation}
with
\begin{equation}
\kappa\equiv \sqrt{k^2-\rho\omega^2/\mu(\omega)},
\qquad
\mu(\omega)\equiv E-i\omega\eta ,
\end{equation}
then, from (\ref{BC.1}) and (\ref{BC.2}), we obtain
the equation to determine the dispersion relation,
\begin{equation}
\left( 2k^2-{\rho\omega^2\over\mu(\omega)}\right)^2
-\gamma{\rho\omega^2k^3\over\mu(\omega)^2}
=
4 k^3\sqrt{k^2-{\rho\omega^2\over\mu(\omega)}}.
\label{det}
\end{equation}
Note that the branch with the positive real part should be taken for
the square roots in order that the mode should be localized around
the surface.

It is easy to examine that this equation gives the Rayleigh wave
dispersion $\omega=c_R\sqrt{E/\rho}\cdot k$ with $c_R=0.9553$,
for the case of the elastic material ($E\ne 0$, $\gamma=\eta=0$)
and the capillary wave dispersion
$\omega=\sqrt{\gamma/\rho}\cdot k^{3/2}$
in the case of the fluid ($\gamma\ne 0$, $E=\eta=0$).

In order to see how the Rayleigh mode and capillary mode exclude each
other,
we consider first the case without the dissipation
($\eta=0$, $E\neq 0$ and $\gamma\neq 0$),
then eq.(\ref{det}) becomes
\begin{equation}
(2-c^2)^2 - c^2\left({\gamma\over E}\right) k = 4\sqrt{1-c^2};
\quad
c\equiv \sqrt{\rho\over E}\,{\omega\over k}.
\end{equation}
The dispersion relation is obtained by determining $c$ for a given 
value of $k$.
Again, note that the square root should have a positive real part.

This equation gives the Rayleigh wave like dispersion
for $k\leq E/\gamma$.
In the case of $k>E/\gamma$, however,
it does not have a solution in the physical branch of the square root.
This means that there is no surface mode in the region of $k> E/\gamma$,
where the capillary stress dominates.

In order to examine what is happening in the region without the surface
modes, we solve the equation of motion
(\ref{t-dep}) numerically with $\eta=0$ under the boundary conditions
(\ref{BC.1}) and (\ref{BC.2}) from the initial state given by $f(z,t=0)$
localized around the surface within a length $\lambda$ as,
\begin{equation}
f(z,t=0) = f_0 e^{-z/\lambda}.
\label{initial}
\end{equation}
The time developments are shown in Fig.1, where energy density,
\begin{equation}
u_e = \sum_{i=x,z}{1\over 2}\rho\dot u_i^2 +
\sum_{i,j=x,z}{1\over 4}E\left({\partial u_i\over\partial x_j}+
{\partial u_j\over\partial x_i}\right)^2,
\end{equation}
is plotted as a function of $z$ for $k=$ 0.5, 1, and 5 $(E/\gamma)$
with the initial configuration with $\lambda=$ 1, 0.5, and 0.5
$(\gamma/E)$, respectively.
Note that the initial shape of $u_e$ is not proportional to
$(e^{-z/\lambda})^2$ because the longitudinal component given by 
(\ref{l.comp}) and (\ref{pw.sol.phi}) decays as $e^{-kz}$ in $z$.

It can be seen that the surface oscillation remains after the pulse is 
gone into the bulk for the case of $k=$ 0.5 $(E/\gamma)$, where the
surface mode is allowed.
On the other hand, the oscillation at the surface is strongly
suppressed for
$k=$ 5 $(E/\gamma)$, where the surface mode is not allowed.
It is interesting to see, in this case, that there is a peak of
oscillation just beneath the surface in addition to the pulse that
travels into the bulk, and the peak of the oscillation diffuses away
as the pulse goes in the bulk.
\end{multicols}
\begin{figure}
\begin{center}
\vspace*{-3.7cm}
\epsfig{file=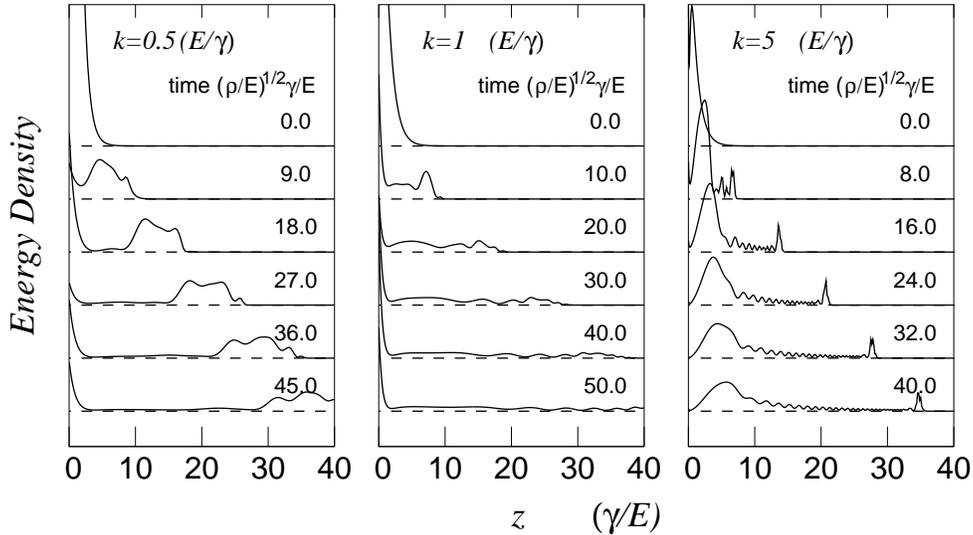, angle=-90, width=0.5\linewidth}
\vspace*{-0.3cm}
\caption{
Time developments of the oscillation for the $\eta=0$ case.
The energy density is plotted against $z$ in the unit of $(\gamma/E)$
for the wave numbers in the $x$ direction $k=$0.5, 1, and 5 $(E/\gamma)$
with the initial state given by (\ref{initial})
with $\lambda =$ 1, 0.5, and 0.5 $(\gamma/E)$, respectively.
The time is shown in unit of $(\rho/E)^{1/2}\gamma/E$.
}
\end{center}
\end{figure}
\begin{multicols}{2}

If the viscosity is introduced into the system, the situation becomes
slightly more complicated.
There takes place the competition among three stresses;
the elastic stress, the surface tension, and the viscous stress.
Depending upon which stress dominates, the mode character is expected
to be one of the followings, the Rayleigh mode, the capillary mode,
or the overdamped liquid mode.
In addition to these, however,
there is the parameter region where no surface mode is allowed,
as we have seen.

\narrowtext
\begin{figure}
\begin{center}
\epsfig{file=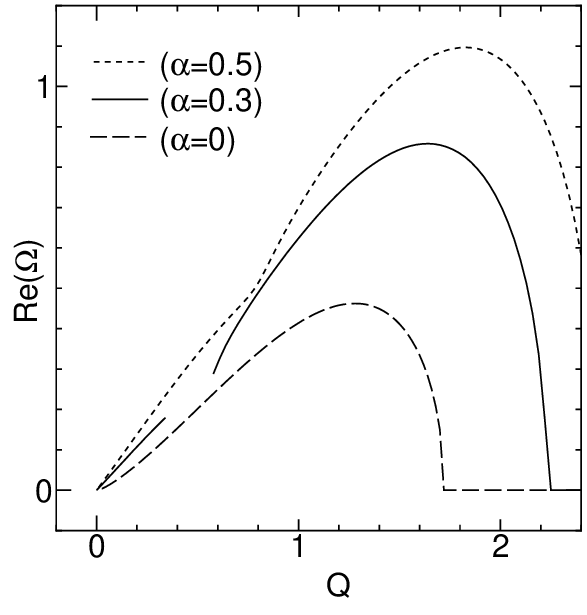, angle=0, width=0.5\linewidth}
\caption{
Dispersion relations for $\alpha\equiv E (\eta/\gamma)^2/\rho=$
0.5, 0.3, and 0.
The real part of $\Omega\equiv (\eta^3/\rho\gamma^2)\omega$ is plotted 
against $Q\equiv (\eta^2/\rho/\gamma)k$.
}
\end{center}
\end{figure}
In Fig.2, the result of dispersion relation for
$E =$ 0.5,  0.3, and 0 $\rho(\gamma/\eta)^2$ is given;
the real part of $\Omega\equiv (\eta^3/\rho\gamma^2)\omega$ is plotted
as a function of $Q\equiv (\eta^2/\rho\gamma)k$.
In the case of $E=$ 0.3 $\rho(\gamma/\eta)^2$,
for small $Q$, the dispersion is linear and the mode is the
Rayleigh wave like, but for an intermediate value of $Q$, the mode becomes
capillary wave like, and for large $Q$, $\Omega$ becomes pure
imaginary and the mode is the overdamped liquid mode.
Between the elastic mode and the capillary mode, there exists 
the region where there exists no surface mode, namely,
there is no solution for
$\Omega$ on the physical branch of the square root in eq.(\ref{det});
we call it the ``gap region''.

The phase diagram in the $E-k$ plane has been obtained numerically by
calculating the dispersion from (\ref{det}) \cite{kn97}, but
the result is illustrated in Fig.3 schematically.
It can be understood as in the following way.
The boundary between the Rayleigh mode and the overdamped liquid mode
is determined by the competition between the elastic stress and the
viscous stress, namely, $Ek|\vec u|$ and $\eta\omega k|\vec u|$,
therefore the boundary is given by
\begin{equation}
k\sim {1\over\eta}\sqrt{\rho E},
\end{equation}
because $\omega\sim\sqrt{E/\rho}\cdot k$ in the Rayleigh mode region.
Similarly, the boundary between the capillary wave and the overdamped
liquid mode is given by
\begin{equation}
k\sim{\rho\gamma/\eta^2}.
\end{equation}
\begin{figure}
\begin{center}
\vspace*{-0.3cm}
\epsfig{file=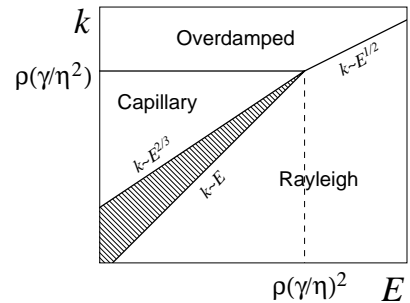, angle=0, width=0.7\linewidth}
\vspace*{-0.3cm}
\caption{
Schematic phase diagram in the $E-k$ plane in the logarithmic scale.
The shaded region is the gap region,
where the surface wave does not exist.
}
\end{center}
\end{figure}

It is instructive to examine how the gap region arises between the
Rayleigh and the capillary region.
For small enough $k$, the elastic stress dominates and the surface
wave is Rayleigh like.
For $E<\rho(\gamma/\eta)^2$, upon increasing $k$,
the surface tension becomes dominant at
\begin{equation}
k\sim E/\gamma ,
\end{equation}
where the surface mode disappears as we have seen for the
dissipationless case.
As $k$ increases further, the elastic stress is taken over by the
viscous stress at
\begin{equation}
k\sim\left(\sqrt{\rho\over\gamma}\,{E\over\eta}\right)^{2/3},
\end{equation}
then the capillary wave emerges \cite{com1}.

It is very suggestive that the existence of the capillary wave depends 
upon the competition between the elastic stress and {\em the viscous
stress}, not the surface tension.
No matter how large the surface tension is, the capillary wave is not
allowed under the existence of elastic stress
unless there exists the viscous stress.
This can be explained as in the following:
The mode characters of the two are too different to form a mixed mode,
therefore, the Rayleigh wave dominates rather than mixes with the
capillary mode when the elastic stress is larger.
On the other hand,
the capillary wave cannot overtake the Rayleigh wave
even when the surface tension at surface is larger than the elastic
stress because the surface tension operates only at the surface
and the displacement field inside is governed by the elastic stress in
the bulk.
The viscosity, however, gives the capillary wave a chance to win by
masking the effects of the elastic stress.


The overdamped liquid mode under the existence of the surface tension
also deserves some comments:
In the bulk, it has the dispersion
\begin{equation}
\omega = -i\, {\eta\over \rho}\, k^2.
\end{equation}
As for the surface mode, there is a similar overdamped mode with the
dispersion
\begin{equation}
\omega=-i\,{\eta\over\rho}\, c_R k^2
\end{equation}
with the same constant $c_R$ with the Rayleigh wave,
and this is the only one in the case without the surface tension.
With the surface tension, however, there appears another mode, whose
damping rate is smaller than the other's;
the dispersion is given by
\begin{equation}
\omega \approx -i\,{\gamma\over 2\eta}\, k
\end{equation}
for $k\gg \gamma\rho/\eta^2$.


Let us examine the experimental relevance of this gap region of the
surface wave.
In the recent experiment\cite{okhk97}, it was reported that the
surface wave shows anomalous
behavior in the tungstic acid; sharp increase and discontinuous drop
of wave speed and attenuation peak of the surface wave in the
frequency range $\sim 100 Hz$
have been observed around the sol-gel transition point, where the
elasticity is very small and the
Rayleigh wave speed is of order of 0.1 $m/s$.
Within the present framework,
the sharp increase of the wave speed can be interpreted as the result of
the group speed increase at the upper edge of the gap in the
dispersion curve.
The discontinuous drop of the wave speed and
the large attenuation of the surface wave come from the suppression of 
the surface oscillation in the gap region.

The typical values of the parameters for the material are
$\rho\sim 10^3 kg/m^3$, $\eta\sim 10^{-2}Pa\cdot s$,
$\gamma\sim 10^{-1}N/m$.
The elastic modulus changes rather rapidly around the sol-gel
transition point, but $E\sim 10 Pa$ in the region where the sound
speed is $0.1 m/s$, then
the expected gap region is $k\approx 10^2\sim 10^{3.3} m^{-1}$, or
$\omega\approx 10^1\sim 10^{2.3} s^{-1}$.
This is consistent with the region where
the anomaly observed.

Before concluding, let us discuss possible effects of structural
relaxation, which we ignore in the present work.
The structural relaxation will relax part of the elastic stress for the
longer time scale than its relaxation time $\tau_{\rm st}$, and can be
expressed by the Maxwell model of visco-elasticity in a
phenomenological level.
If we include the effect, the capillary mode region would extend into the
gap region to some degree in the small $k$ side because the structural
relaxation reduces the bulk elastic stress in longer time scale than the
structural relaxation time $\tau_{\rm st}$.
The way how the gap region shrinks from the present result (Fig.3)
depends upon detailed structure of the model and the relevant time
scale;
for $\omega \gg 1/\tau_{\rm st}$ ($\omega \ll 1/\tau_{\rm st}$),
the part of the stress can be treated as elastic (viscous).
The basic feature of the phase diagram, however, should not change
as long as there exists a pure elastic component of the stress
in the system as we assumed in the present study.

In summary, we found that, in the visco-elastic material, there is a
parameter region where the surface mode is absent and it corresponds to
the region where the surface wave anomaly is found in the recent
experiment on tungstic acid\cite{okhk97}.

The authors would like to thank H.Okabe, K.Kuboyama, K.Hara, and S.Kai
for showing their experimental data prior to publication.
This work is partially supported by Grant-in-Aid for Scientific
Research (C) (\# 0940468) provided by the Ministry of
Education, Science, Sports, and Culture, Japan.


\end{multicols}

\begin{references}
\bibitem{com1}
The gap region is missing in the phase diagram Fig.1 of \cite{hpp91},
where the boundary between the capillary mode and the elastic mode is
drawn.
\end{references}
\end{document}